\documentclass[sigconf,edbt]{acmart-edbt2023}

\usepackage{multirow}

\setcopyright{rightsretained}

\acmDOI{}

\acmISBN{978-3-89318-088-2}
\acmConference[EDBT 2023]{26th International Conference on Extending Database Technology (EDBT)}{28th March-31st March, 2023}{Ioannina, Greece}
\acmYear{2023}

\begin{document}

\title{Implementing and Evaluating E2LSH on Storage}

\author{Yu Nakanishi, Kazuhiro Hiwada, Yosuke Bando, Tomoya Suzuki,\\
        Hirotsugu Kajihara, Shintaro Sano, Tatsuro Endo, Tatsuo Shiozawa}
\email{
  {yu.nakanishi,kazuhiro.hiwada,yosuke1.bando,tomoya.suzuki,hirotsugu.kajihara,shintarou.sano,tatsuro1.endo,tatsuo.shiozawa}@kioxia.com}
\affiliation{%
  \institution{Kioxia Corporation}
  \country{Japan}
}

\begin{abstract}
  Locality sensitive hashing (LSH) is one of the widely-used approaches to approximate nearest neighbor search (ANNS) in high-dimensional spaces.
  The first work on LSH for the Euclidean distance, E2LSH, showed how ANNS can be solved efficiently at a sublinear query time in the database size with theoretically-guaranteed accuracy, although it required a large hash index size.
  Since then, several LSH variants having much smaller index sizes have been proposed. 
Their query time is linear or superlinear, but they have been shown to run effectively faster because they require fewer I/Os when the index is stored on hard disk drives and because they also permit in-memory execution with modern DRAM capacity.

  In this paper, we show that E2LSH is regaining the advantage in query speed with the advent of modern flash storage devices
  such as solid-state drives (SSDs). 
  We evaluate E2LSH on a modern single-node computing environment and analyze its computational cost and I/O cost, from which we derive storage performance requirements for its external memory execution. 
  Our analysis indicates that E2LSH on a single consumer-grade SSD can run faster than the state-of-the-art small-index methods executed in-memory.
  It also indicates that E2LSH with emerging high-performance storage devices and interfaces can approach in-memory E2LSH speeds.
  We implement a simple adaptation of E2LSH to external memory, E2LSH-on-Storage (E2LSHoS), and evaluate it for practical large datasets of up to one billion objects using different combinations of modern storage devices and interfaces.
  We demonstrate that our E2LSHoS implementation runs much faster than small-index methods and can approach in-memory E2LSH speeds, and also that its query time scales sublinearly with the database size beyond the index size limit of in-memory E2LSH.

  
\end{abstract}

\maketitle

\section{Introduction}

Nearest neighbor search (NNS) in high-dimensional Euclidean spaces is an important operation in many diverse application domains,
such as databases, text search, multimedia indexing, pattern recognition, and machine learning \cite{SurveyNNS,SurveyANNS,PLSH,MLLSH,VIDEOLSH,AUDIOLSH}.
NNS in high-dimensional spaces is known to be computationally expensive due to the curse of dimensionality \cite{Curse1,Curse2}.
The common approach to overcoming the high computational cost is to design efficient approximate nearest neighbor search (ANNS) algorithms,
which find neighbors that are close enough to the query item instead of the exact nearest neighbors.
Locality sensitive hashing (LSH) is one of the widely used approximate methods
{among others such as graph-based and tree-based approaches \cite{DATASET,ANNBenchmarks}}.
The advantages of LSH are its theoretically guaranteed accuracy and practical speed \cite{LSH,SurveyLSH}.
The first work on LSH in the Euclidean space, introduced by Datar et al. \cite{E2LSH} and later named E2LSH \cite{E2LSHpackage},
showed how ANNS can be solved efficiently with sublinear query time in the number $n$ of objects in the database.
On the other hand, a large hash index of size superlinear in $n$ has to be maintained,
which has limited its applicability to relatively small databases.
Although one billion ANNS was demonstrated using a 100-node cluster \cite{PLSH}, in order to handle such large databases on a single node,
E2LSH would have required external memory for storing a large hash index.
Hence, several LSH variants have been proposed to reduce the index size \cite{C2LSH,QALSH,I-LSH,SRS,PM-LSH,R2LSH}, which we collectively refer to as small-index LSH methods.
Although they come at the cost of sublinear query time, meaning their computational load is linear or superlinear,
their good empirical performance has made them constitute the mainstream of Euclidean LSH in the recent literature \cite{DATASET}.

In this paper, we deviate from this trend and show that E2LSH on a single node for various practical datasets, including billion-class large databases, can be made faster than small-index methods through
the effective use of modern flash storage devices such as solid-state drives (SSDs).
These devices have dramatically shorter access times than hard disk drives (HDDs)
while providing terabytes of capacity necessary to hold large hash indices,
suggesting that E2LSH may regain its competitiveness through external memory implementation (which we call E2LSH-on-Storage or E2LSHoS for short).
To examine when this turning point happens, we evaluate the E2LSH algorithm on a modern single-node computing environment and analyze its computational cost and I/O cost.
Based on these numbers, we identify the storage performance requirements in order for E2LSHoS to run at a given speed.
Our analysis indicates that external memory needs to provide a random read performance of a few hundred kIOPS (I/O operations per second) in order to compete with the state-of-the-art small-index methods SRS \cite{SRS} and QALSH \cite{QALSH} executed in-memory. 
While such an IOPS number is hard to come by with HDDs, 
it can easily be achieved by a single consumer-grade NVMe\texttrademark{} SSD if the drive receives multiple read requests in parallel to take advantage of the multiple flash memory dies inside the drive.
This means that E2LSHoS can be faster than the small-index methods if we can adapt the algorithm
using asynchronous I/Os.
By using the same analysis framework, we can also study whether E2LSHoS can be as fast as in-memory E2LSH.
The storage requirements turn out to be a random read performance of a few MIOPS and a CPU I/O overhead of tens of nanoseconds.
This suggests that in-memory-class speeds are achievable using emerging high-performance storage devices and I/O interfaces. 

To verify our observations, we implement E2LSHoS incorporating the external memory adaptations including asynchronous reads.
We conduct experiments using several combinations of storage devices and interfaces,
and show that E2LSHoS indeed runs faster than the small-index methods even with a single commodity NVMe\texttrademark{} SSD providing 350 kIOPS.
By placing a large index on storage, the runtime memory (DRAM) usage of E2LSHoS is comparable to the small-index methods.
We also show that E2LSHoS approaches in-memory speeds by using high IOPS drives and a lightweight I/O interface with a small CPU overhead.
Having in-memory-class speeds effectively amounts to relaxing the index size limit of E2LSH imposed by DRAM without compromising its speed or accuracy.
One can enjoy the benefit of sublinear query time of E2LSH (i.e., more speedup gains over linear time methods as the database size $n$ increases)
for larger databases that were previously possible only on a multi-node cluster.
We experimentally validate the sublinearity of E2LSHoS query time up to one billion objects. 

In summary, our contributions are as follows.
(1) We experimentally analyze the computational and I/O cost of the E2LSH algorithm on a modern single-node hardware environment in order to identify the storage performance requirements for its external memory implementation.
(2) We implement E2LSHoS using asynchronous I/Os to maximize the random read performance of storage, and evaluate it for practical large datasets of up to one billion objects using different combinations of modern storage devices and interfaces on a single node.
(3) We demonstrate the strengths of E2LSHoS: it runs much faster than small-index methods 
using commodity NVMe\texttrademark{} SSDs; 
it can approach in-memory E2LSH speeds using emerging storage devices and interfaces;
and its query time scales sublinearly with the database size beyond the index size limit of in-memory E2LSH.
These results suggest that larger-index LSH methods are once again becoming increasingly worth exploring. 
%

\section{Preliminaries}

This section reviews ANNS, LSH, and E2LSH to provide the background information for our analysis and evaluation. 

\subsection{Approximate Nearest Neighbor Search}

Let $ \mathcal{D} $ be a database of $n$ objects, each represented by a point in the $d$-dimensional space, 
where the dissimilarity between two objects $o_1$ and $o_2$ is measured by their Euclidean distance $||o_1, o_2||$.
%
Given a query $q \in \mathbb{R}^d$ and an approximation ratio $c \; (\ge 1 )$,
$c$-approximate nearest neighbor search (ANNS) returns an object $o \in \mathcal{D}$ satisfying $||q,o|| \le c \cdot ||q,o^*||$, where $o^* \in \mathcal{D}$ is the exact nearest neighbor.
Likewise, top-$k$ $c$-ANNS returns a set of objects $o_i \in \mathcal{D} \; (1 \le i \le k) $ satisfying $||q,o_i|| \le c \cdot ||q,o^*_i|| $, where $o^*_i$ is the $i$-th nearest neighbor.

\subsection{Locality Sensitive Hashing}

Locality sensitive hashing (LSH) is one of the most widely used ANNS techniques,
whose two main advantages are theoretical guarantees and empirical performance.
It uses hashes under which two objects collide with a higher probability when they are closer, allowing one to quickly collect candidate neighbors before checking their actual distances to the query.
Such a hash function
that maps a point $o \in \mathbb{R}^d$ to an integer hash value can be constructed as \cite{E2LSH},

\begin{equation}
\label{eqn:LSH_function}
h(o) = \left\lfloor \frac{a \cdot o + b}{w} \right\rfloor,
\end{equation}
where
$a \in \mathbb{R}^d$ is a random vector whose $d$ elements are
drawn from the standard normal distribution $\mathcal{N}(0,1)$,
$w$ is a bucket width, and $b$ is a uniform random number on $[0, w)$.
Intuitively, the vector dot product $a \cdot o$ projects the object $o$ onto a randomly-oriented line $a$,
and the division by $w$ followed by the floor operation splits the line into segments of length $w$, forming hash buckets.
Shifting by $b$ randomizes the bucket boundaries.
With Equation~\ref{eqn:LSH_function}, two objects $o_1$ and $o_2$ are likely to fall in the same bucket when their distance $s = ||o_1,o_2||$ is small.
%
That is, the probability $Pr[h(o_1) = h(o_2)]$ of hash collision is a monotonically decreasing function of the distance $s$ \cite{E2LSH}.
We denote this function by $p_w(s)$ in what follows.
The subscript indicates that it also depends on the bucket width $w$.

\subsection{E2LSH}
\label{sec:E2LSH}

\newcommand{\maxnumsearch}{S}

Here we explain the method proposed by Datar et al. that solves $c$-ANNS efficiently using LSH \cite{E2LSH}, which we adapt to external memory. 
It is commonly referred to as E2LSH after the software package based on it \cite{E2LSHpackage}, although they have some differences.
We use the original algorithm \cite{E2LSH}, which has a query time guarantee. 

Consider the subproblem called $(R, c)$-near neighbor (NN) search, where, given a fixed radius $R$, the task is to determine whether there is an object within distance $R$ from the query $q$.
And if there is one, we report any object within distance $cR$. If not, we report so.
By successively performing $(R, c)$-NN for increasing radii $R = 1, c, c^2, c^3, \cdots$ until some object is reported, the reported object is a solution to $c^2$-ANNS
\cite{LSH, LSBFOREST}.

In order to solve $(R, c)$-NN, LSH functions of Equation~\ref{eqn:LSH_function} can be used to collect candidate neighbors because we have
\begin{align}
Pr[h(q) = h(o)] &\ge p_1 &\mathrm{for} \;\; ||q,o|| &\le R \\
Pr[h(q) = h(o)] &\le p_2 &\mathrm{for} \;\; ||q,o|| &\ge cR
\end{align}
with $p_1 > p_2$
thanks to the monotonically decreasing collision probability $p_w(||q,o||)$. 
Near objects within distance $R$ collide with the query at a high probability of at least $p_1 = p_w(R)$, whereas far objects beyond distance $cR$ collide at a low probability of at most $p_2 = p_w(cR)$.
E2LSH uses a series of \emph{compound hashes} each consisting of multiple LSH functions as
\begin{equation}
\label{eqn:compound_hash}
g_i(o) = (h_{i1}(o), h_{i2}(o), \cdots, h_{im}(o)),
\end{equation}
where $h_{ij}$ is the $j$-th LSH function in the $i$-th compound hash.
Each compound hash consists of $m$ hashes with randomly chosen $a$ and $b$, and we consider all of these $m$ hash values to be combined together to point to one hash bucket.
This more effectively rejects far objects because the collision probability will be raised to the $m$-th power as $Pr[g_i(q) = g_i(o)] \le p_2^m \ll 1$.
As this also decreases the collision probability for near objects to a lesser extent, multiple ($L$) compound hashes are used so at least one of them catches near objects.

In preprocessing, E2LSH constructs $L$ separate sets of hash buckets.
For each object $o \in \mathcal{D}$, it calculates compound hash values $g_1(o), g_2(o), \cdots, g_L(o)$, and adds the object $o$ to the $L$ buckets they point to.
The mapping from a compound LSH hash value to the corresponding bucket is maintained using a table or a standard hash, called a \emph{hash table}.
In the query phase, E2LSH computes the hash values of the query $q$ as $g_1(q), g_2(q), \cdots, g_L(q)$, and searches all of those $L$ buckets.
Objects in these buckets are candidate neighbors.
We compute their actual distances to the query, and report those that are within $cR$. As the number of candidate objects may be large, E2LSH stops the search after examining $\maxnumsearch$ candidates.


%
By adjusting the parameters $m$, $L$, and $\maxnumsearch$, we can control the probability of successfully finding the $(R, c)$-NN solutions.
With
\begin{equation}
\label{eqn:E2LSHparams}
m = \log_{1/p_2} n, \quad
L = n^\rho, \quad
\maxnumsearch = 2L
\end{equation}
where $\rho = \log(1/p_1)/\log(1/p_2) < 1$,
the success probability becomes $1/2 - 1/e$ \cite{E2LSH}.

Since E2LSH calculates $mL$ hash values for a given query and computes the distances of up to $\maxnumsearch$ objects, the time complexity is $O(dmL + d\maxnumsearch) = O(d n^\rho \log n)$, which is sublinear in the database size $n$.
On the other hand, it consumes large space. In addition to the $O(dn)$ database size, each object is stored in $L$ hash buckets, resulting in $O(nL) = O(n^{1+\rho})$ superlinear hash index size. 

In performing $(R, c)$-NN for increasing radii,
the maximum radius $R_{\max}$ we need to look at is 
given as $R_{\max} = 2 x_{\max} \sqrt{ d }$, where $x_{\max}$ is the maximum absolute value of the object coordinates.
The number $r$ of radii to be searched is thus $r = \lceil \log_c R_{\max}\rceil$.
This depends on the extent of the object coordinates, but not on the database size. 

\subsection{Other LSH methods}
\label{sec:otherLSH}

Many methods have been proposed to address the large index size of E2LSH.
Multi-Probe LSH collects more candidate neighbors by probing multiple nearby hash buckets,
resulting in near linear index sizes \cite{MULTIPROBELSH}.
It builds upon the entropy-based theory \cite{EntropyLSH} while improving on the practical performance, but its theoretical guarantees are yet to be established \cite{MultiProbeLSHRevisited}.
LSB-Forest converts $m$ hash values into a Z-order value and indexes it using a B-tree \cite{LSBFOREST}.
Different lengths of leading bits of the Z-order values correspond to different search radii, eliminating the need for preparing hash indices for different radii, although this limits the approximation ratio $c$ to an integer.
The query time and index size remain sublinear and superlinear, respectively. 
C2LSH introduced \emph{collision counting} and \emph{virtual rehashing} techniques to achieve $O(n \log n)$ query time and index space \cite{C2LSH}. 
Albeit still superlinear, this is a significant reduction in the index size, but the query time also became superlinear.
Subsequent work further improved performance using query-aware bucketing (QALSH \cite{QALSH}) and query-centered incremental search (I-LSH \cite{I-LSH}), with the same 
space and time complexities.
Other methods use LSH to project objects onto a low-dimensional space, where they collect candidate neighbors using an R-tree (SRS \cite{SRS}) and a PM-tree (PM-LSH \cite{PM-LSH}), which leads to linear space and time complexities.
R2LSH \cite{R2LSH} combines two LSH functions to construct multiple 2D projections, and candidates are searched within query-centric balls.
The space and time complexities are both linear.

Although the query time of these small-index methods is either linear or superlinear, the reduced I/O cost for external memory has had a large impact, allowing them to run faster than large-index methods.
Moreover, even if they initially assumed external memory execution, 
the typical DRAM capacity has increased since then, permitting even faster in-memory execution.
However, the storage performance has also increased in the meantime, likely more so than DRAM.
We are interested in seeing if large-index methods can be faster again even if they still need to resort to external memory execution, for which we take the seminal sublinear time E2LSH as a first-step example.
The idea of running E2LSH-like methods on external memory dates back to 1999 \cite{LSHonExternalMemory}, and this paper is by no means new in that respect.
{However, previous HDD-based studies had to deal with costly I/Os. There has been an attempt to minimize I/Os aggressively by using only a small number of hashes and avoiding distance checking \cite{NVTree}, but this deviates from the original E2LSH formulation, compromising its theoretical guarantees.}
With the advent of modern storage devices, {reports on their use in ANNS 
are emerging \cite{SSDANN1,SSDANN2}.
Our intention is to push this trend} and provide insights into the impact of storage on the E2LSH query time, which we hope will help the community to further explore this direction.



\subsection{Other ANNS Methods}
\label{sec:otherANNS}

{
While the focus of this paper is LSH, there are many other approaches to ANNS \cite{SurveyNNS,SurveyANNS}.
Recent benchmarks \cite{DATASET,ANNBenchmarks} compare methods from three major categories among others: graph-, tree-, and LSH-based.
Graph-based methods \cite{NSG,HNSW,KGRAPH,DiskANN} construct graphs whose vertices representing objects are connected by edges when they are close to each other, so that NNs can be searched efficiently by traversing the edges.
Tree-based methods partition the search space into a hierarchy using tree structures such as randomized $k$-d trees \cite{FLANN} and random projection trees \cite{ANNOY}.
Neither graph- nor tree-based method has theoretical guarantees, but they are considered to generally achieve higher empirical speeds \cite{DATASET,ANNBenchmarks}.
The major advantages of LSH over these approaches include theoretical guarantees 
as well as relatively simple index structures based on hashing, which are easy to maintain and update.
We believe our work is a step forward toward
accelerating the LSH-based approach
without losing its theoretical footing, so that it will continue to be one of ANNS algorithms from which practitioners can choose. 
}

\section{Experimental Setup}
\label{sec:setup}

This section explains the experimental setup used both for analyzing the E2LSH algorithm in Sec.~\ref{sec:analysis} and for evaluating its external memory implementation, E2LSHoS, in Sec.~\ref{sec:results}.
We focus on the case where the hash index (both hash tables and buckets) is placed on external memory while the database itself is kept on DRAM, as the hash index is by far the most space-consuming data structure. 

\subsection{LSH Methods}
We use three LSH methods, E2LSH, SRS \cite{SRS}, and QALSH \cite{QALSH}.
E2LSH has in-memory and external memory versions.
In-memory E2LSH is implemented by modifying the E2LSH package \cite{E2LSHpackage} so it performs $c$-ANNS.
The external memory version, E2LSHoS, will be presented in Sec.~\ref{sec:method}.
To compare E2LSH(oS) with small-index methods, we select SRS and QALSH as representative benchmarks.
They are among the state-of-the-art LSH-based methods with quality guarantees, and are used as benchmarks by previous reports \cite{DATASET,I-LSH,PM-LSH,R2LSH}.
{Since their index size is small, we use their in-memory implementations \cite{SRSSRC,QALSHMem} posted by the respective authors of the methods, and thus the entire index is stored in the main memory.
While more recent methods such as PM-LSH \cite{PM-LSH} and I-LSH \cite{I-LSH} are published, they do not come with open implementations.
PM-LSH has the same linear time complexity as SRS and the paper reports 30\% faster query than SRS.
I-LSH has the same superlinear time complexity as QALSH. It reduces I/Os at the cost of more computation \cite{SurveyLSH} and thus in the in-memory settings, it is likely to perform similarly to QALSH.
As they should perform similarly to SRS and QALSH, respectively, we believe we can provide more grounded evaluation using the open implementations of SRS and QALSH.}

\subsection{Evaluation Metrics}

We compare methods by their query times required to achieve the same level of accuracy.
We often present \emph{speedup} gains by taking the query time ratio to show how many times one method is faster than the other.
We use the \emph{overall ratio} metric for accuracy.
For top-$k$ ANNS,
it is defined as
$\frac{1}{k} \sum_{i=1}^{k} ||o_i, q|| / ||o_i^*, q||$,
where $ o_i $ is the $i$-th ANN returned by a given method, and $ o_i^* $ is the exact $i$-th NN. 
A smaller value means higher accuracy, It becomes 1 if exact NNs are returned.
An overall ratio of 1.05 is used as our default target.



\subsection{Parameter Settings}
\label{sec:parameters}

Different methods have different algorithmic parameters that affect their performance.
%
The E2LSH parameters $(m, L, \maxnumsearch)$ are calculated by Equation~\ref{eqn:E2LSHparams}. 
Given the approximation ratio $c$ (we use $c=2$), 
the value of $\rho$ can be adjusted by varying the bucket width $w$.
When $\rho$ is small, both of the index size $O(n^{1+\rho})$ and query time $O(d n^\rho \log n)$ become smaller, 
but this generally leads to lower empirical accuracy. 
Therefore, we set $\rho$ large enough 
to achieve the desired range of accuracy,
and fine tune the accuracy by introducing a scaling parameter $\gamma$ to modify $m$ such that $m = \gamma \log_{1/p_2}n$.
This scaling is useful as it does not affect the index size once $\rho$, hence $L$, is fixed for a given dataset.
The scaling also modifies the success probability, but that can be compensated for by the choice of $\maxnumsearch$, and it does not change the sublinear query time complexity as long as $\gamma > \rho$. 

For SRS, we set $c=4$ (equivalent to $c=2$ in E2LSH) and the success probability to $1/2 - 1/e$.
SRS projects objects in $\mathcal{D}$ onto a low dimensional space.
After experimenting with various values for the projection dimension (denoted by $m$ in \cite{SRS}), we found $m=8$ works well for all of our evaluations.
We control the accuracy by varying the maximum number of data points 
to be checked (denoted by $T'$).

For QALSH, we set the success probability to $1/2 - 1/e$. For lack of other tweakable parameters, we adjust the accuracy through $c$.

\subsection{Datasets}
\label{sec:datasets}

We use eight widely-used datasets in Table \ref{table:Datasets}, 
covering a broad range of data types. 
Relative Contrast (RC) \cite{RC} and Local Intrinsic Dimensionality (LID) \cite{LID} are shown as proxies for the hardness of the datasets.
Smaller RC and larger LID imply harder datasets.
We conduct experiments using the queries 
accompanying each dataset.

\begin{table}[h]
  \caption{Datasets}
  \label{table:Datasets}
 \centering
  \begin{tabular}{lrrcrrl}
   \hline
   Name & $n$ ($\times 10^3$) & $d$ & Data & RC & LID & Type\\
   \hline
   {MSONG}          & 983 &   420 & float & 4.04 &  23.8 & {Audio} \\
   {SIFT}         & 1,000 &   128 & byte  & 3.20 &  21.7 & {Image} \\
   {GIST}         & 1,000 &   960 & float & 2.14 &  47.3 & {Image} \\
   {RAND}         & 1,000 &   100 & float & 1.42 &  49.6 & {Synthetic} \\
   {GLOVE}        & 1,183 &   100 & float & 2.20 &  22.1 & {Text} \\
   {GAUSS}        & 2,000 &   512 & float & 1.14 & 147.1 & {Synthetic} \\
   {MNIST}        & 8,000 &   784 & byte  & 3.00 &  20.4 & {Image} \\
   {BIGANN}       & 1,000,000 &   128 & byte  & 3.55 &  25.4 & {Image} \\
   \hline
  \end{tabular}
\end{table}

\subsection{Execution Environment}
\label{sec:environment}
Our experimental machine is equipped with two Intel\textregistered{} Xeon\textregistered{} Gold 5218 2.3GHz 32-core CPUs and DDR4 DRAM 768 GB (32 GB $\times$ 24), running Linux (Ubuntu 18.04 and kernel 5.3.0).
For all the LSH methods we compare, hash value and distance computations are accelerated using AVX-512.
Unless otherwise noted,
we use a single core to preclude performance variability due to multi-core execution.
The reason for the high capacity DRAM is to be able to run in-memory E2LSH for as large databases as possible.
It allows us to deal with up to 100 million objects, 
although it falls far short of the index size for one billion objects.
By using external memory, we do not need this DRAM capacity: with BIGANN dataset, E2LSHoS consumes 140 GB in total, of which 130 GB is the database size.


\subsection{Storage Devices}
\label{sec:devices}

As external memory for E2LSH, we use three types of 
flash-based storage devices:
consumer-grade NVMe\texttrademark{} SSDs (cSSD), enterprise-grade NVMe\texttrademark{} SSDs (eSSDs), and prototype device XLFDDs \cite{XLDD0}.
Their measured random read performances are shown in Table~\ref{table:ssd_performance}. 
These IOPS numbers play a key role as hash buckets will be read randomly.
Flash storage offers orders of magnitude higher IOPS than HDDs, especially when I/O requests are fed in parallel and the number of concurrent I/Os being processed, called the \emph{queue depth}, is large.
%
eSSDs are equipped with low-latency flash memory chips and provide higher IOPS than cSSDs.
The reason why we include XLFDDs in our evaluation is not because of their even higher IOPS,
but of their lightweight I/O interface as explained next.

\begin{table}[h]
  \caption{Storage devices and their random read performance}
  \label{table:ssd_performance}
  \begin{center}
  \begin{tabular}{|c|c|c|c|}
    \hline
    & & \multicolumn{2}{|c|}{kIOPS${}^\dagger$} \\
    \cline{3-4}
    Type & Storage model & \multicolumn{2}{|c|}{Queue depth} \\
    \cline{3-4}
    & &  1 & 128 \\
    \hline
    \multirow{2}{*}{cSSD} & KIOXIA BiCS FLASH\texttrademark{} XG5, & \multirow{2}{*}{7.2} & \multirow{2}{*}{273} \\
    & 2 TB, NVMe\texttrademark{} 1.4, PCIe\textregistered{} 3.0 & & \\
    \hline
    \multirow{2}{*}{eSSD} & KIOXIA XL-FLASH\texttrademark{} FL6, & \multirow{2}{*}{27.6} & \multirow{2}{*}{1,400} \\
    & 800 GB, NVMe\texttrademark{} 1.4, PCIe\textregistered{} 4.0${}^\ast$ & & \\
    \hline
    \multirow{2}{*}{XLFDD} & KIOXIA XL-FLASH\texttrademark{} Demo Drive & \multirow{2}{*}{132.3} & \multirow{2}{*}{3,860} \\
    & 520 GB, PCIe\textregistered{} 3.0 & & \\
    \hline
    \multirow{2}{*}{HDD${}^\ddagger$} & Seagate IronWolf & \multirow{2}{*}{0.21} & \multirow{2}{*}{0.54} \\
    & 10 TB, 7200rpm && \\
    \hline
  \end{tabular}
  \end{center}
  \raggedright
  \footnotesize{
    ~ * Our evaluation is limited to the performance of PCIe\textregistered{} 3.0 supported by the CPU. \\
    \dag \, Measured at 512 bytes rather than 4 kB in order not to be bandwidth-limited. \\
    \ddag \, Listed here for reference only. 
   }
\end{table}

\subsection{Storage Access Interfaces}

Issuing I/O requests to storage consumes some CPU time.
This overhead can have a non-negligible impact on the query time of E2LSHoS issuing a large number of I/Os. 
The CPU overhead depends on the storage access interfaces, and conventional interfaces often consume a lot of CPU time going through system calls and interruption handlers to access external devices.

Recently, more lightweight interfaces are becoming available including io\_uring \cite{IOURING} and SPDK \cite{SPDK}.
In addition, XLFDD is designed to permit a more lightweight interface than NVMe\texttrademark{} SSDs \cite{XLDD1}.
We use these three interfaces in our evaluation.
Table~\ref{table:interfaces} shows their measured CPU overheads
expressed in terms of the time a CPU core spends in issuing one I/O request, as well as its reciprocal indicating the maximum IOPS one core may be able to draw from storage.
These numbers depend on our execution environment and may be different from benchmarks reported elsewhere.
Our aim is to assess the impact of the CPU overhead on E2LSHoS and not to decide which interface is better: faster versions are already emerging. 

\begin{table}[h]
  \caption{Storage interfaces and their CPU overhead}
  \label{table:interfaces}
  \centering
  \begin{tabular}{lrr}
    \hline
    Interface       & CPU time per I/O    & Max IOPS/core \\
    \hline
    io\_uring (version 2.0)  & 1.0 $\mu$sec  & 1.0 MIOPS \\
    SPDK (version 21.10)  & 350 nsec      & 2.9 MIOPS \\
    XLFDD interface                           &  50 nsec      &  20 MIOPS \\
    \hline
  \end{tabular}
\end{table}

\section{Experimental Analysis of E2LSH}
\label{sec:analysis}

In this section, we experimentally analyze the characteristics of the E2LSH algorithm.
The purposes of the study is to identify the storage performance requirements for its external memory implementation, E2LSHoS, to do the following:
\begin{itemize}
\item{To run faster than small-index LSH methods SRS and QALSH}
\item{To approach in-memory E2LSH speeds}
\end{itemize}

For clarity and simplicity, we first analyze the case of top-1 ANNS ($k=1$), and then examine the influence of $k \; (>1)$ in top-$k$ ANNS.

Note that we do \emph{not} use storage devices for any of the experiments in this section.
We run in-memory E2LSH to characterize the algorithm without relying on particular storage devices or external memory implementations.
The analysis in this section guides our external memory implementation in Sec.~\ref{sec:method}, whose performance will be evaluated in Sec.~\ref{sec:results} using actual storage devices.

\subsection{Query Time Model of E2LSHoS}
\label{sec:model}

\newcommand{\EELSHtime}{T_{\rm E2LSH}}
\newcommand{\SRStime}{T_{\rm SRS}}
\newcommand{\QALSHtime}{T_{\rm QALSH}}
\newcommand{\EELSHoStimeSync}{T_{\rm E2LSHoS, sync}}
\newcommand{\EELSHoStimeAsync}{T_{\rm E2LSHoS, async}}
\newcommand{\EELSHtimeCompute}{T_{\rm compute}}
\newcommand{\EELSHtimeAccess}{T_{\rm access}}
\newcommand{\CPUIOoverhead}{T_{\rm request}}
\newcommand{\storageIOtime}{T_{\rm read}}
\newcommand{\DRAMtime}{T_{\rm load}}
\newcommand{\IOcount}{N_{\rm I/O}}
\newcommand{\IOcountArg}[1]{N_{\rm I/O, #1}}
\newcommand{\targetTime}{T_{\rm target}}

We use query time models to analyze the E2LSHoS speed without relying on actual implementations.
We consider decomposing the query time into a computational part (hash value calculation and distance checking) and I/O part (hash bucket reads).
{Note that distance checking does not require I/Os as we keep object coordinates on DRAM as described in Sec.~\ref{sec:setup}.}

\begin{figure}[thb]
  \begin{center}
  \includegraphics[width=\columnwidth]{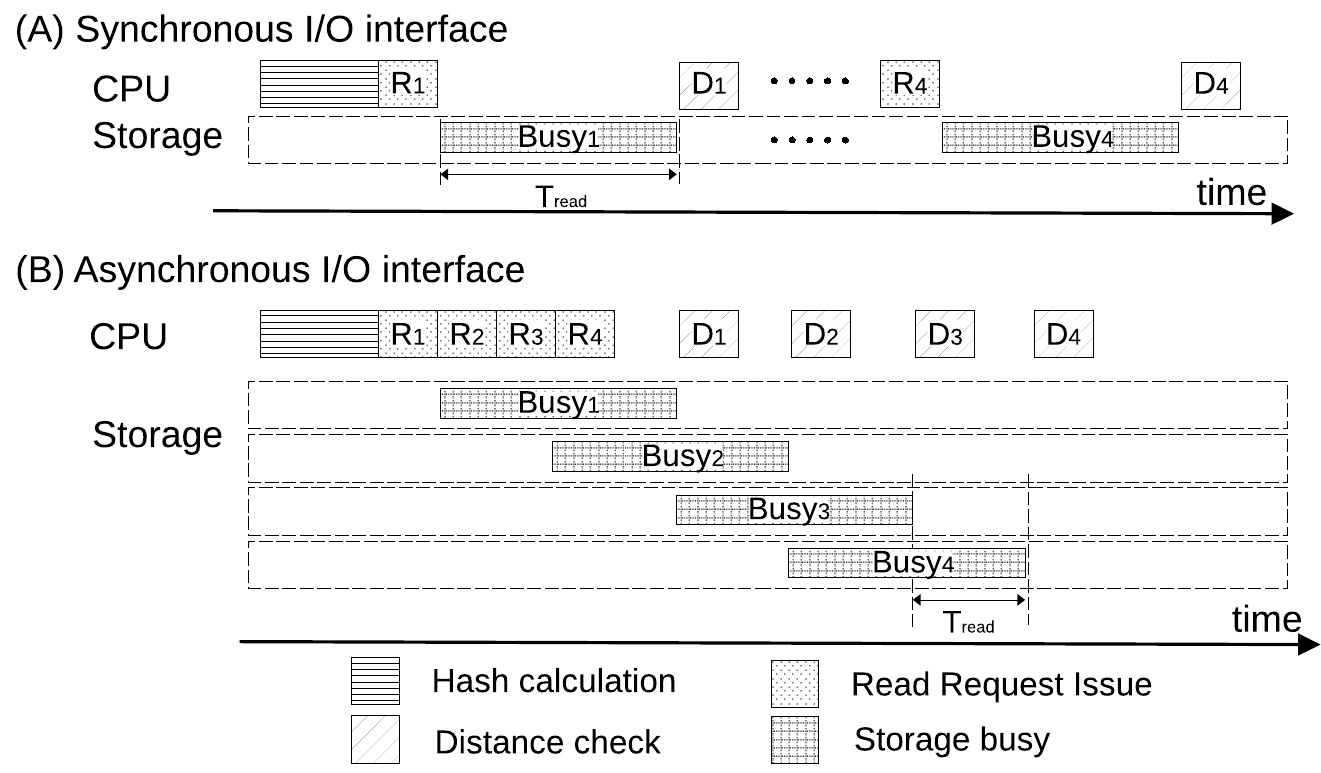}
  \caption{Query time models of E2LSHoS}
  \label{fig:io_interface}
  \end{center}
\end{figure}

We first consider an adaptation to external memory using a simpler, synchronous I/O interface as shown in Figure~\ref{fig:io_interface}(A).
To fetch buckets, the CPU issues an I/O (read) request and waits for the requested data to arrive from the storage before proceeding with a next operation. 
The query time is modeled as follows.
\begin{equation}
\label{eqn:E2LSHoStimeSync}
\EELSHoStimeSync = \EELSHtimeCompute + \IOcount \cdot (\CPUIOoverhead + \storageIOtime),
\end{equation}
where $\EELSHtimeCompute$ is the total time spent for computation (of hash values and distances), $\IOcount$ is the number of I/Os (four in the figure), $\CPUIOoverhead$ is the CPU overhead time per I/O request, and $\storageIOtime$ is the time for the storage to return requested data (i.e., latency).
Note that the lengths of these operations are not to scale in the figure.

As the synchronous implementation leaves both the CPU and storage idle for some time, one can minimize the idle times by using an asynchronous I/O interface as shown in Figure~\ref{fig:io_interface}(B),
such that the CPU issues I/O requests for multiple buckets without blocking,
allowing the storage to process requests in parallel. 
The figure shows only one unit of processing from hash calculation to distance checking 
corresponding to one search radius for one query, but multiple radii and queries
can be interleaved, 
{so that the computation and I/O times will mostly overlap.}
What we are interested here is the average processing time per query, which is determined by the longer of the CPU time and the storage read time.
Hence, the query time can be expressed as
\begin{equation}
\label{eqn:E2LSHoStimeAsync}
\EELSHoStimeAsync = \max\{\EELSHtimeCompute + \IOcount \cdot \CPUIOoverhead, \; \IOcount \cdot \storageIOtime \}.
\end{equation}
{While it is unlikely that the computation and I/O times overlap perfectly in practice, this simplified cost model is useful for making our analysis tractable.}
%
Note that, in the asynchronous case, $\storageIOtime$ is shorter than the storage latency, 
as shown in the figure.
In either case, $\storageIOtime$ represents the time the storage spends per I/O.
In other words, its reciprocal $\storageIOtime^{-1}$ is the random read performance of the storage device measured in IOPS, as buckets are accessed in no particular order.
The performance depends on the queue depth, the number of requests that are fed in parallel, as mentioned in Sec.~\ref{sec:devices}.
A queue depth of 1 corresponds to the synchronous case, whereas it is over 1 in the asynchronous case, leading to better performance.

The dependencies of the variables in Equations~\ref{eqn:E2LSHoStimeSync} and \ref{eqn:E2LSHoStimeAsync} are:
\begin{itemize}
\item $\EELSHtimeCompute$ depends on the dataset, query, and machine.
\item $\IOcount$ depends on the dataset and query. 
\item $\CPUIOoverhead$ depends on the storage interface and machine.
\item $\storageIOtime$ depends on the storage device and queue depth.
\end{itemize}
Using the setup described in Sec.~\ref{sec:setup}, we experimentally evaluate the computational cost $\EELSHtimeCompute$ and the number $\IOcount$ of I/Os of E2LSH in Secs.~\ref{sec:computational_cost} and \ref{sec:io_cost}, respectively.
Based on these numbers, we can identify the storage requirements $\CPUIOoverhead$ and $\storageIOtime$ for the E2LSHoS query time to reach a given value.

Let $\targetTime$ denote the target query time.
In order for E2LSHoS to reach it, we need to have $\EELSHoStimeSync \le \targetTime$ in the synchronous case.
By rearranging Equation~\ref{eqn:E2LSHoStimeSync}, we obtain
\begin{equation}
\CPUIOoverhead + \storageIOtime \le \frac{\targetTime - \EELSHtimeCompute}{\IOcount}.
\end{equation}
In the synchronous case, the latency $\storageIOtime$ dominates, being at least tens of microseconds with typical SSDs while the CPU overhead $\CPUIOoverhead$ is at most 1 $\mu$sec as in Table~\ref{table:interfaces}.
Therefore, we omit $\CPUIOoverhead$ from the inequality 
and take the reciprocal as
\begin{equation}
\label{eqn:iops_sync}
\storageIOtime^{-1} \ge \frac{\IOcount}{\targetTime - \EELSHtimeCompute}.
\end{equation}
This inequality provides the required IOPS value for the random read performance $\storageIOtime^{-1}$ of the storage.

In the asynchronous case, from $\EELSHoStimeAsync \le \targetTime$ and Equation~\ref{eqn:E2LSHoStimeAsync}, we can derive
\begin{align}
\label{eqn:iops_async_cpu}
\CPUIOoverhead^{-1} &\ge \frac{\IOcount}{\targetTime - \EELSHtimeCompute}, \\
\label{eqn:iops_async_storage}
\storageIOtime^{-1} &\ge \frac{\IOcount}{\targetTime}. 
\end{align}
The first inequality provides the requirement for the CPU overhead $\CPUIOoverhead^{-1}$, 
while the second inequality determines the required random read performance $\storageIOtime^{-1}$ of the storage.
In addition to the fact that random read performance of flash storage is orders of magnitude higher with asynchronous I/Os as shown in Table~\ref{table:ssd_performance},
the IOPS requirement in Equation~\ref{eqn:iops_async_storage} is relaxed by having a larger denominator than the synchronous case in Equation~\ref{eqn:iops_sync}.
Therefore, we will use Equations~\ref{eqn:iops_async_cpu} and \ref{eqn:iops_async_storage} to identify the storage performance needed for E2LSHoS to run faster than small-index methods in Sec.~\ref{sec:srs_speed} and to approach in-memory E2LSH speeds in Sec.~\ref{sec:in_memory_speed}.

\subsection{Computational Cost of E2LSH}
\label{sec:computational_cost}

First, we evaluate the computational cost $\EELSHtimeCompute$ of the E2LSH algorithm and reconfirm that it is significantly smaller than that of SRS and QALSH.
We run all of the three algorithms in-memory, and hence there is no storage I/O:
we can simply measure the query time of each method to evaluate its computational cost. 

\begin{figure}[t]
  \begin{center}
  \includegraphics[scale=0.45]{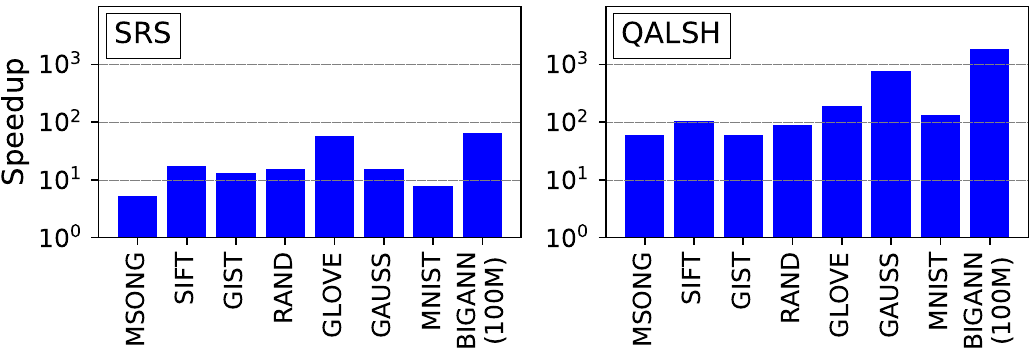}
  \vspace*{-3mm}
  \caption{Speedup gains of E2LSH over SRS and QALSH
  }
  \label{fig:e2lsh_vs_srs}
  \end{center}
\end{figure}

The speedup gains of E2LSH over SRS and QALSH are shown in Figure~\ref{fig:e2lsh_vs_srs}.
We had to take a subset (100 million objects) of BIGANN dataset for E2LSH to run in-memory due to the DRAM capacity limit as described in Sec.~\ref{sec:environment}.
The speedup is consistently well over 1. 
In many cases, E2LSH is faster by an order of magnitude or two.

\noindent
{\bf Observation 1.}
The computational cost of E2LSH is generally much less expensive than that of SRS and QALSH. 

In addition, we found that SRS is consistently faster than QALSH.
Thus, we use SRS as the sole baseline representing small-index methods in what follows.
{The speed differences between E2LSH, SRS, and QALSH can be explained by their query time complexities (sublinear, linear, and superlinear, respectively).
To gain further insight, we examined the behaviors of E2LSH and SRS and found that, on average in our experiments, E2LSH checkes hundreds of hash buckets to find thousands of candidates to answer one query, whereas SRS visits tens of thousands of R-tree nodes to find thousands of candidates. Hence, the speed difference primarily comes from the number of search index elements (buckets or tree nodes) they need to visit before finding candidates.}

\subsection{I/O Cost of E2LSH}
\label{sec:io_cost}

Next, we analyze the I/O cost of E2LSH.
As we treat the storage-related variables $\CPUIOoverhead$ and $\storageIOtime$ as unknowns in Equations~\ref{eqn:E2LSHoStimeSync} and \ref{eqn:E2LSHoStimeAsync}, what we analyze here is the average number $\IOcount$ of I/Os required to answer a query. 

To do this, we recall that the E2LSH algorithm reads $L$ buckets for each search radius and repeats this up to $r$ times for increasing radii until an answer is found.
Thus, E2LSHoS will read up to $Lr$ buckets per query, but oftentimes the search ends before exhausting all the radii.
By running in-memory E2LSH, we can experimentally calculate the average number $\bar{r}$ of radii (averaged over queries) to be searched for each dataset, which is summarized in Table~\ref{table:num_of_hash}. 

\begin{table}[htb]
  \caption{Average number of hash bucket reads per query}
  \label{table:num_of_hash}
 \centering
 \begin{tabular}{c}
   \begin{tabular}{lrrrr}
     \hline
     \multirow{2}{*}{Dataset} & \# hashes & Total \#   & Avg. \#          & Avg. \# I/Os          \\
                              & $L$       & radii $r$  & radii $\bar{r}$  & $\IOcountArg{\infty}$ \\
     \hline
     MSONG        & 16 & 11 & 5.76 & 133.6 \\
     SIFT         & 25 & 11 & 9.08 & 347.5 \\
     GIST         & 32 & 4 & 1.70 & 48.7 \\
     RAND         & 48 & 4 & 3.00 & 196.5\\
     GLOVE        & 51 & 5 & 3.82 & 317.2 \\
     GAUSS        & 19 & 8 & 6.00 & 190.8\\
     MNIST        & 18 & 13 & 11.60 & 393.7\\
     BIGANN(100M) & 48 & 11 & 9.03 & 791.0 \\
     \hline
   \end{tabular}
 \end{tabular}
\end{table}

In order to calculate the number of I/Os required to read $L\bar{r}$ buckets,
we need to take the following two points into account.
Firstly, E2LSHoS will also need to issue I/Os for hash table access.
As explained in Sec.~\ref{sec:E2LSH}, E2LSH maintains a hash table that maps a compound LSH hash value to the corresponding bucket.
In external memory implementation, buckets will be pointed to by their storage addresses.
As we need a hash table on the order of the database size $n$ for each of the $Lr$ hashes, the total data size will be $O(Lrn)$,
which can be large in itself. 
Hence we assume the hash tables are also on external memory, and one I/O is required to read the storage address of a bucket from a hash table before accessing the bucket itself.
Secondly, E2LSHoS may need to issue multiple I/Os to read one bucket.
We note that each bucket can contain an arbitrary number of objects as it depends on the spatial distribution of objects in the database and hash collision probabilities.
Moreover, E2LSH stops reading the content of a bucket in the middle if the maximum number $\maxnumsearch$ of candidates is reached.
In order to save the stroage bandwidth in reading a variable portion of a variable-sized bucket, a relatively small read block size is desirable.
With a finite block size $B$, the number of I/Os per bucket may be more than one, in addition to the one I/O for the hash table.

\begin{figure}[t]
  \begin{center}
  \includegraphics[width=0.95\columnwidth]{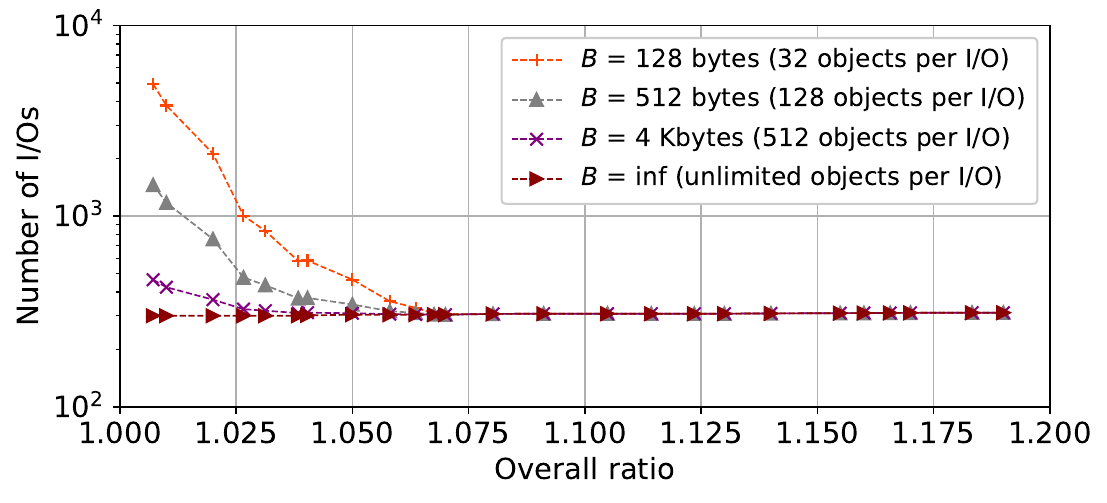}
  \vspace*{-3mm}
  \caption{Average number of I/Os required to answer a query in SIFT dataset for varying block size $B$}
  \label{fig:sift_io_block_sizes}
  \end{center}
\end{figure}

We first make a conservative estimate of the I/O count by assuming that every bucket fits in the block size $B$, effectively ignoring the second point above. 
The rightmost column of Table~\ref{table:num_of_hash} shows the minimum number $\IOcountArg{\infty}$ of I/Os thus calculated.
The subscript of $\infty$ denotes the block size is assumed to be arbitrarily large.
Note that empty buckets are not counted as it is easy to avoid issuing I/Os for them, and hence the number is smaller than $2L\bar{r}$.
As we adjust the ANNS accuracy without changing $L$ as described in Sec.~\ref{sec:parameters}, the number $\IOcountArg{\infty}$ does not depend on the accuracy.

To make a more practical estimate of the I/O count, we examine how the number of I/Os changes depending on the accuracy (affecting the number of objects in a bucket and the candidate number limit $\maxnumsearch$) and choice of read block size $B$.
Figure~\ref{fig:sift_io_block_sizes} shows the case of SIFT dataset.
Here we set the size of each object to 4 bytes as this has enough bits to distinguish billions of objects.
As shown, the IO count tends to be larger for higher accuracy (smaller overall ratio) as buckets tend to contain more objects.
Naturally, more IOs are required with smaller block sizes.
Note, however, that the choice of block size does not affect the sublinear query time of E2LSHoS, as the I/O count is capped at the maximum number $\maxnumsearch$ of objects to be checked, which grows sublinearly in the database size $n$.

\noindent
{\bf Observation 2.}
E2LSH requires at least several hundred I/Os per query for many workloads.
The I/O count becomes larger for higher accuracy and smaller block sizes.




\subsection{Requirements for SRS Speeds}
\label{sec:srs_speed}

\begin{figure}[t]
  \begin{center}
  \includegraphics[width=0.95\columnwidth]{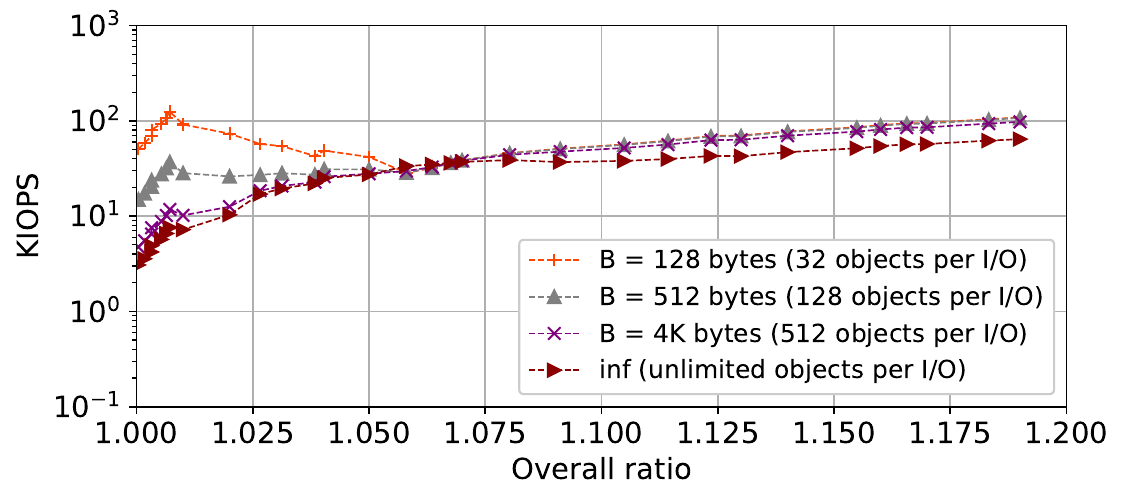}
  \vspace*{-3mm}
  \caption{IOPS requirements for SRS speeds with varying block size $B$ for SIFT dataset}
  \label{fig:latency_async_block_sizes}
  \end{center}
\vspace{6mm}
  \begin{center}
  \includegraphics[width=0.95\columnwidth]{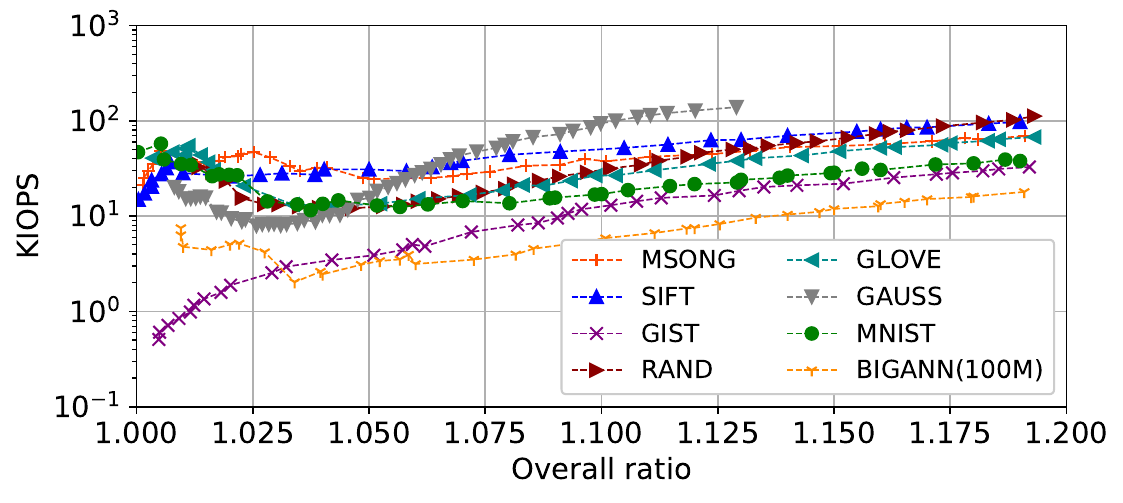}
  \vspace*{-3mm}
  \caption{IOPS requirements for SRS speeds with a block size of $B = 512$ bytes
}
  \label{fig:latency_async_512B}
  \end{center}
\end{figure}

Now that we have the values for $\EELSHtimeCompute$ and $\IOcount$, we can solve for the storage requirements.
%
%
%
For asynchronous E2LSHoS to achieve a comparable speed to the query time $\SRStime$ of in-memory SRS, we set $\targetTime = \SRStime$ in Equations~\ref{eqn:iops_async_cpu} and \ref{eqn:iops_async_storage} as
\begin{align}
\label{eqn:iops_async_cpu_srs}
\CPUIOoverhead^{-1} &\ge \frac{\IOcount}{\SRStime - \EELSHtimeCompute} \\
\label{eqn:iops_async_storage_srs}
\storageIOtime^{-1} &\ge \frac{\IOcount}{\SRStime} 
\end{align}

We first take a look at the second inequality providing the random read performance requirement.
We begin by examining the effect of block size $B$ in Figure~\ref{fig:latency_async_block_sizes}, where we plot $\IOcount/\SRStime$, the right-hand side of Equation~\ref{eqn:iops_async_storage_srs}, for SIFT dataset using the I/O counts $\IOcount$ in Figure~\ref{fig:sift_io_block_sizes}.
The minimum required IOPS value with $B = \infty$ decreases as it moves toward higher accuracy (to the left) because the SRS query time $\SRStime$ increases while the I/O count $\IOcount$ of E2LSHoS stays the same as in Figure~\ref{fig:sift_io_block_sizes}.
With finite block sizes, the required IOPS increases in the high accuracy region because the I/O counts increase there.
Still, even with a small block size of $B = 128$ bytes, the required IOPS does not exceed that 
of the low accuracy region.
Since increasing the block size will consume more bandwidth, we consider $B = 512$ bytes a reasonable choice as it is the minimum size supported by typical NVMe\texttrademark{} SSDs.
With the choice of $B = 512$ bytes, Figure~\ref{fig:latency_async_512B} plots IOPS requirements for all the datasets.
We can see that a few hundred kIOPS is required in order to satisfy Equation~\ref{eqn:iops_async_storage_srs} for all the cases (datasets and accuracy levels).

\begin{figure}[t]
  \begin{center}
  \includegraphics[width=0.95\columnwidth]{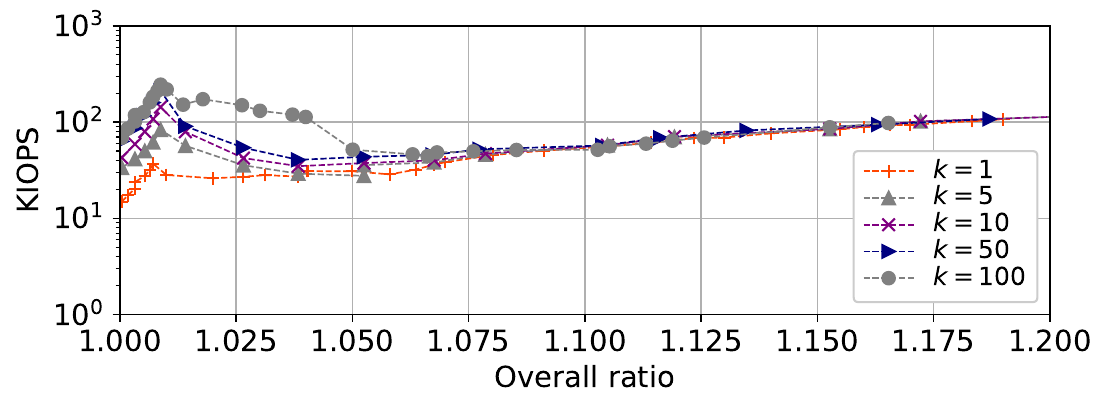}
  \vspace*{-3mm}
  \caption{IOPS requirements for SRS speeds for varying $k$ for SIFT dataset}
  \label{fig:latency_async_512B_sift_topk}
  \end{center}
\end{figure}

This IOPS value is expected to be sufficient for a larger database size $n$ and larger $k$ in top-$k$ ANNS.
When the database size $n$ becomes larger, the IOPS requirement tends to be lower.
This is because the numerator $\IOcount$ of Equation~\ref{eqn:iops_async_storage_srs} grows sublinearly while the denominator $\SRStime$ grows linearly as SRS is a linear time algorithm.
The requirement is thus proportional to $n^{\rho-1}$, which decreases with $n$.
It is a moderate decrease, though, as ten-fold increase in $n$ will not lead to one-tenth the IOPS requirement.
Hence, looking at the BIGANN(100M) line in Figure~\ref{fig:latency_async_512B}, one-billion databases will still likely require up to ten kIOPS.
With top-$k$ ANNS,
we observe higher IOPS requirements for larger $k$ in the high accuracy region as shown in
Figure~\ref{fig:latency_async_512B_sift_topk}, which is still not significantly higher than the requirement in the low accuracy region at $k=1$.


Next, we examine Equation~\ref{eqn:iops_async_cpu_srs} providing the CPU overhead requirement.
Its right-hand side is larger than Equation~\ref{eqn:iops_async_storage_srs}, but only so by a small margin since $\SRStime \gg \EELSHtimeCompute$ as shown in Sec.~\ref{sec:computational_cost}, and the requirement is still a few hundred kIOPS (in terms of maximum IOPS/core).
That is, the CPU overhead time $\CPUIOoverhead$ must be less than a few microseconds.
This requirement 
always holds as long as we use storage interfaces shown in Table~\ref{table:interfaces}.







Referring to Table~\ref{table:ssd_performance}, a few hundred kIOPS is hard to achieve with HDDs as hundreds of them would be required, 
while it is easily overcome by a single cSSD with asynchronous I/Os.
A cSSD with synchronous I/Os falls far short (and recall that the IOPS requirement is larger in the synchronous case as described in Sec.~\ref{sec:model}).

\noindent
{\bf Observation 3.}
The random read performance required for storage devices to allow E2LSHoS to achieve comparable query speeds to SRS is a few hundred kIOPS.
This IOPS number can be obtained by using a single cSSD with asynchronous implementation.

\subsection{Requirements for In-memory Speeds}
\label{sec:in_memory_speed}

\begin{figure}[t]
  \begin{center}
  \includegraphics[width=0.95\columnwidth]{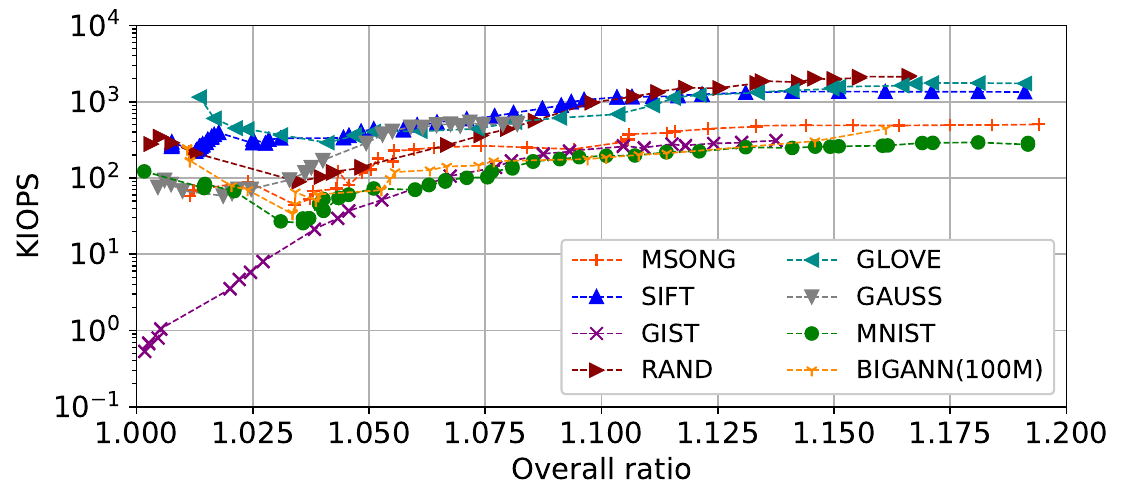}
  \vspace*{-3mm}
  \caption{IOPS requirements for in-memory E2LSH speeds}
  \label{fig:latency_inmem}
  \end{center}
\vspace{6mm}
  \begin{center}
  \includegraphics[width=0.95\columnwidth]{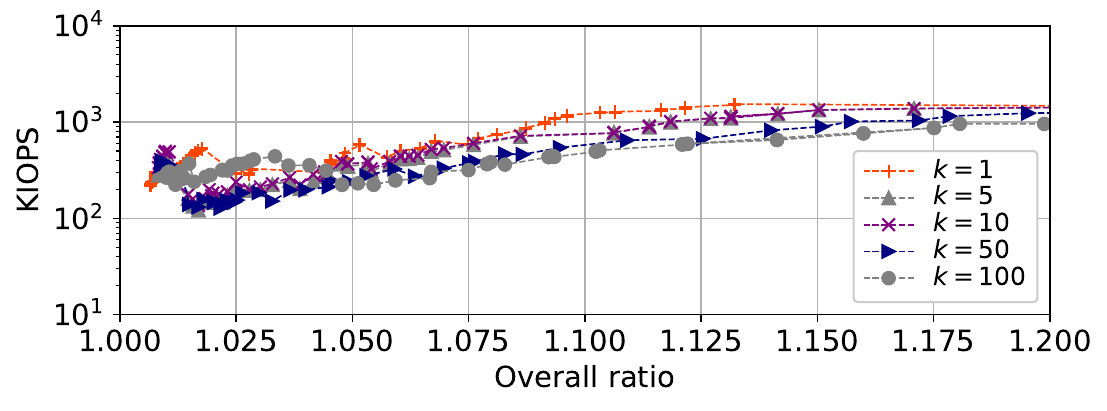}
  \vspace*{-3mm}
  \caption{IOPS requirements for in-memory E2LSH speeds for varying $k$ for SIFT dataset}
  \label{fig:latency_inmem_sift_topk}
  \end{center}
\end{figure}

We turn our attention to whether E2LSHoS can be as fast as in-memory E2LSH.
By setting our target time to the query time of in-memory E2LSH as $\targetTime = \EELSHtime$ in Equations~\ref{eqn:iops_async_cpu}, we have
\begin{align}
\label{eqn:iops_inmem_cpu}
\CPUIOoverhead^{-1} &\ge \frac{\IOcount}{\EELSHtime - \EELSHtimeCompute} \\
\label{eqn:iops_inmem_storage}
\storageIOtime^{-1} &\ge \frac{\IOcount}{\EELSHtime} 
\end{align}
Again, we first look at the second inequality and plot the right-hand side in Figure~\ref{fig:latency_inmem}, using the I/O count $\IOcountArg{512}$ assuming a block size of $B = 512$ bytes.
The plots tell us that the required random read performance to reach in-memory-class speeds for all the cases is a few MIOPS.
Although a single cSSD is no longer sufficient, this IOPS number is easily achievable by using multiple drives 
or by using higher performance drives such as eSSDs and XLFDDs.

This IOPS requirement is expected to hold for a larger database size $n$ and larger $k$ in top-$k$ ANNS.
This is because the in-memory E2LSH time $\EELSHtime$ and its I/O count $\IOcount$, the denominator and numerator of Equation~\ref{eqn:iops_inmem_storage}, both grow in the same way in $n$ and $k$:
we know they both grow sublinearly in $n$, and
while we do not have equations for $k$,
no substantial change in the IOPS requirements is observed for larger $k$ as shown in Figure~\ref{fig:latency_inmem_sift_topk}.

Next, we examine the first inequality in Equation~\ref{eqn:iops_inmem_cpu}.
In Sec.~\ref{sec:computational_cost}, we estimated the computational time $\EELSHtimeCompute$ of E2LSHoS by the query time $\EELSHtime$ of in-memory E2LSH, which has been sufficient as we have only used the fact that $\EELSHtimeCompute \ll \SRStime$ for our analysis so far.
However, to examine Equation~\ref{eqn:iops_inmem_cpu},
we note that $\EELSHtimeCompute$ is slightly shorter than $\EELSHtime$.
As in-memory E2LSH places everything in DRAM including a large hash index, it has a much larger memory footprint than E2LSHoS, leading to a higher likelihood of CPU cache misses.
As in-memory E2LSH alone does not tell us how much CPU time is accounted for by the increased footprint,
we estimate it by running a workload that imitates bucket reading and distance checking of E2LSH.
It reads buckets randomly from within a specified memory footprint and performs a fixed amount of computation.
For each dataset, we compare two cases.
One with the footprint size E2LSH would use, and the other with a footprint that is smaller by the index size.
With the smaller memory footprint, we observe a decrease in the memory stall time using \textit{perf} command.
As a result, the runtime decreases around 10\% for all the datasets.
By plugging $\EELSHtimeCompute = 0.9 \, \EELSHtime$ into Equation~\ref{eqn:iops_inmem_cpu}, we estimate the requirement to be
\begin{equation}
\CPUIOoverhead^{-1} \ge 10 \frac{\IOcount}{\EELSHtime}.
\end{equation}
The right-hand side is a scaled version of the random read performance requirement of a few MIOPS, meaning $\CPUIOoverhead^{-1}$ must be at least a few tens of MIOPS in terms of maximum IOPS/core, or $\CPUIOoverhead$ must be less than a few tens of nanoseconds.
This requirement is in the range of what XLFDD interface provides.

\noindent
{\bf Observation 4.}
In-memory E2LSH speeds are achievable with storage devices having random read performance of a few MIOPS along with a lightweight storage interface incurring only a small CPU overhead time of no more than a few tens of nanoseconds per I/O.
\section{E2LSH-on-Storage Implementation}
\label{sec:method}

This section presents how to implement E2LSH on external memory,
which we call E2LSH-on-Storage (E2LSHoS).

The analysis in Sec.~\ref{sec:analysis} made it clear that asynchronous I/O is crucial in allowing E2LSHoS to run faster than SRS and to approach in-memory speeds.
Many I/O requests need to be issued in parallel in order to increase the queue depth and make full use of the random read performance of storage devices.
In addition, we have learned 
that buckets containing many objects can be read in small blocks without impacting the performance, as long as the storage provides a few MIOPS.
We design E2LSHoS in what follows based on these points as well as on some practical considerations.

{Because the E2LSH algorithm performs random access to a large index space and the access locality is low (this will be demonstrated in Sec.~\ref{sec:additional_experiments}), E2LSHoS does not employ any caching for I/Os:
the OS page cache is bypassed and all the data comes from storage.
}

\subsection{Data Structure}
\label{subsec:datastructure}


We store both hash tables and buckets on external memory.
The data structure of the hash bucket is shown in Figure~\ref{fig:bucket}.
Since the numbers of objects in the buckets vary, we use a linked list similar to \cite{LSHonExternalMemory} whose elements (called \emph{bucket blocks}) each contain a header and multiple object IDs.
The header holds a pointer (address on the storage) to the next block, while an object ID points to the $d$-dimensional coordinates of that object on DRAM.
We use a block size of 512 bytes, the minimum unit for a typical SSD read command.
The header size is 16 bytes, where 8 bytes are used for the link pointer,
2 bytes to store the number of objects in the block as it may be less than the block can hold,
and the rest is a padding (reserved for debug purposes).
For the reason described below in Sec.~\ref{sec:fingerprint}, we attach a \emph{fingerprint} (FP) to an object ID to form a 5-byte \emph{object info} entry.
The space overhead coming from the header and fingerprints reduces the number of objects per block to 99 ($= (512-16)/5$) from 128, but this number is still large enough to
keep the IOPS requirement from growing
according to the analysis in Sec.~\ref{sec:analysis}.
The storage address of the first bucket block is stored in a hash table.

\subsection{Hash Value Size and Fingerprints}
\label{sec:fingerprint}

\newcommand{\hashvaluebits}{v}
\newcommand{\hashtablebits}{u}

\begin{figure}[t]
  \begin{center}
  \includegraphics[width=\columnwidth]{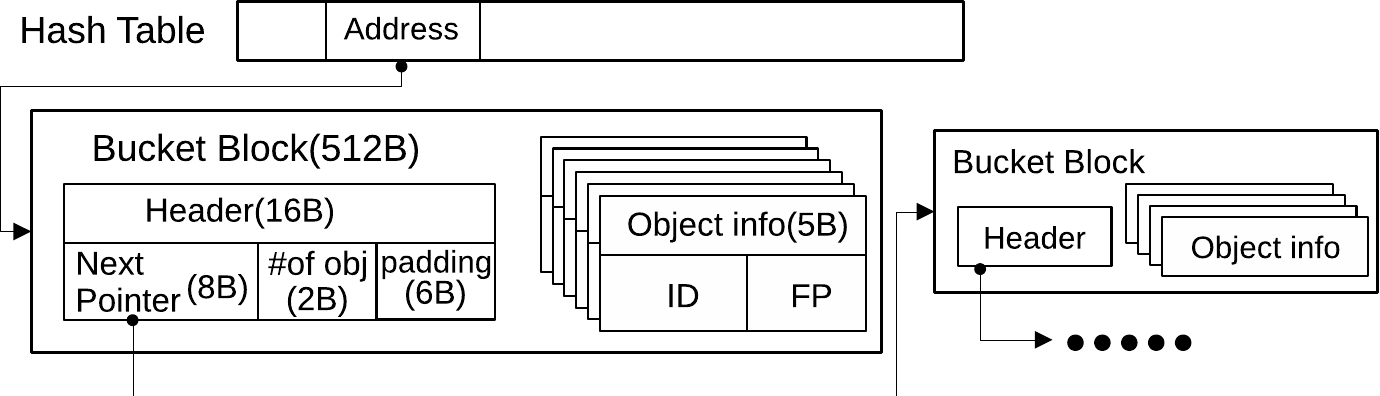}
  \caption{Data structure of the hash bucket}
  \label{fig:bucket}
  \end{center}
\vspace{6mm}
  \begin{center}
  \includegraphics[width=1.0\columnwidth]{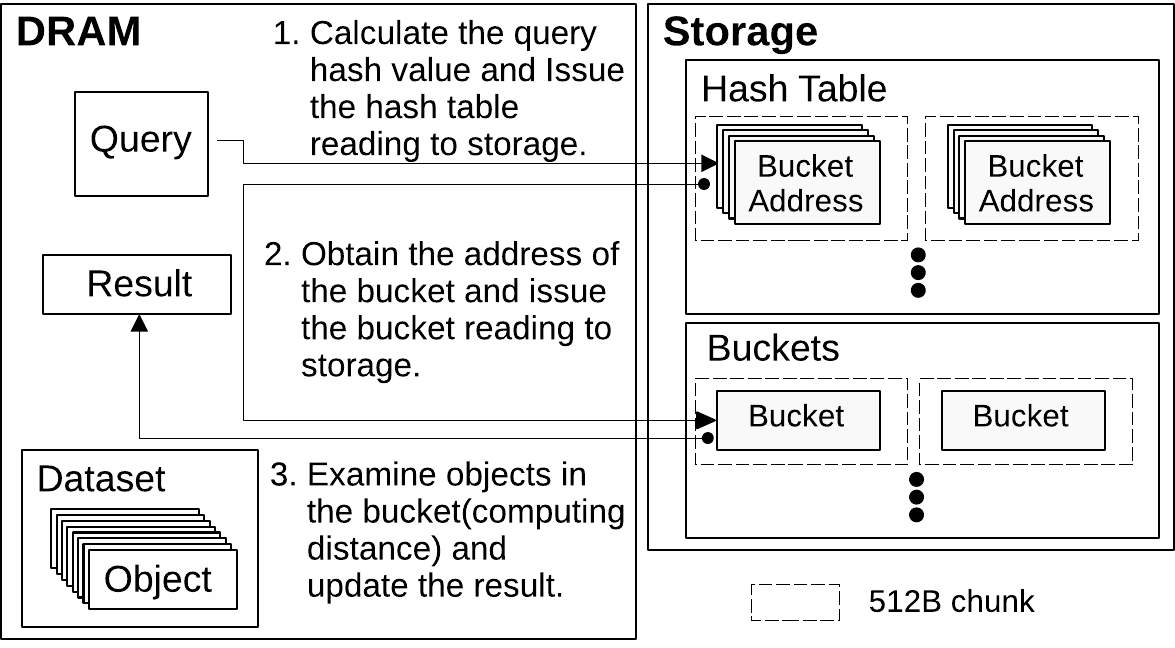}
  \caption{Hash index structure and query processing steps
  }
  \label{fig:overview}
  \end{center}
\end{figure}

Hash values need to be represented by a finite number ($\hashvaluebits$) of bits in practice, which can introduce false hash collisions due to the limited precision.
While a larger number for $\hashvaluebits$ is desirable, it increases the cost of managing hash values.
Our implementation uses $\hashvaluebits = 32$ bits in order for ideal hash functions to be able to distinguish one billion objects.
Meanwhile, we use a hash table that looks at $\hashtablebits$ bits of a hash value, and attach the remaining $\hashvaluebits - \hashtablebits$ bits as a fingerprint to the object info in the bucket.
This means we effectively use $\hashtablebits$-bit hash values at the hash table level.
If we set $\hashtablebits \approx \log_2 n$, the hash table size will be $O(Lrn)$, and the false collision probability will be still low.
We can further reject false collisions coming from the use of $\hashtablebits < \hashvaluebits$ bits by checking the fingerprints when reading hash buckets, ensuring 32-bit precision.
We use $\hashtablebits$ that is slightly smaller than $\log_2 n$ as long as it does not substantially increase false collisions.
Since the object ID can be represented by $\lceil \log_2 n \rceil$ bits, the object info in Figure~\ref{fig:bucket} will use $\lceil \log_2 n \rceil + \hashvaluebits - \hashtablebits$ bits in total, which can be larger than $\hashvaluebits = 32$ bits.
Thus we allocate 5 bytes to the object info.

\subsection{LSH Index Construction}

E2LSHoS constructs a hash index as follows.
For each radius $R \in \{1, c, c^2, \cdots, c^{r-1}\}$ and compound hash $l \in \{1, 2, \cdots, L\}$,
we randomly generate a compound hash function,
calculate the hash values of all the objects $o_j$ in the database $\mathcal{D}$, and add the objects to the buckets corresponding to their hash values.
If the bucket size exceeds a bucket block of 512 bytes, we allocate a new empty bucket block on the storage
and put its address in the header of the current bucket block.
Finally, we write the hash table on the storage.


\subsection{Query Processing}
\label{sec:query_processing}

For each search radius $R$ and compound hash $l$, E2LSHoS finds ANNs of a given query as shown in Figure~\ref{fig:overview}.
We calculate a hash value of the query and issue a read request for the bucket address in the hash table (Step 1).
We then issue a read request for the bucket data using that address (Step 2). 
Upon receiving the bucket data,
we examine the objects in the bucket by computing their distances to the query and update the query result (Step 3).
The bucket may be read in multiple bucket blocks by traversing the link addresses in the headers.
%
In order to issue many I/O requests in parallel to maximize the storage utility, we interleave multiple queries.
We issue read requests for $L$ buckets successively for one query, and switch the context to process another query while waiting for data. 
\section{Evaluation Results}
\label{sec:results}

We evaluate our E2LSHoS implementation presented in Sec.~\ref{sec:method} using various storage configurations (types of device, numbers of devices, and interfaces) and compare it with SRS and in-memory E2LSH.
Table~\ref{table:configurations} shows the device types and numbers.
When using multiple devices, the number is set so that it is sufficient to store the index for BIGANN dataset.
Both cSSD and eSSD may be combined with io\_uring or SPDK interface.
XLFDD uses its own interface.

\begin{table}[h]
  \caption{Storage device configurations}
  \label{table:configurations}
  \centering
  \begin{tabular}{crrr}
    \hline
    Device & Number & Total capacity & Total random read \\
    \hline
    cSSD   & 1   & 2 TB    & 273 kIOPS \\
    cSSD   & 4   & 8 TB    & 1.1 MIOPS \\
    eSSD   & 1   & 800 GB  & 1.4 MIOPS \\
    eSSD   & 8   & 6.4 TB  & 11.2 MIOPS \\
    XLFDD  & 12  & 6.2 TB  & 46.3 MIOPS \\
    \hline
  \end{tabular}
\end{table}

\subsection{Effect of Storage Configurations}
\label{sec:eval_storage_config}

\begin{figure}[t]
  \begin{center}
  \includegraphics[width=\columnwidth]{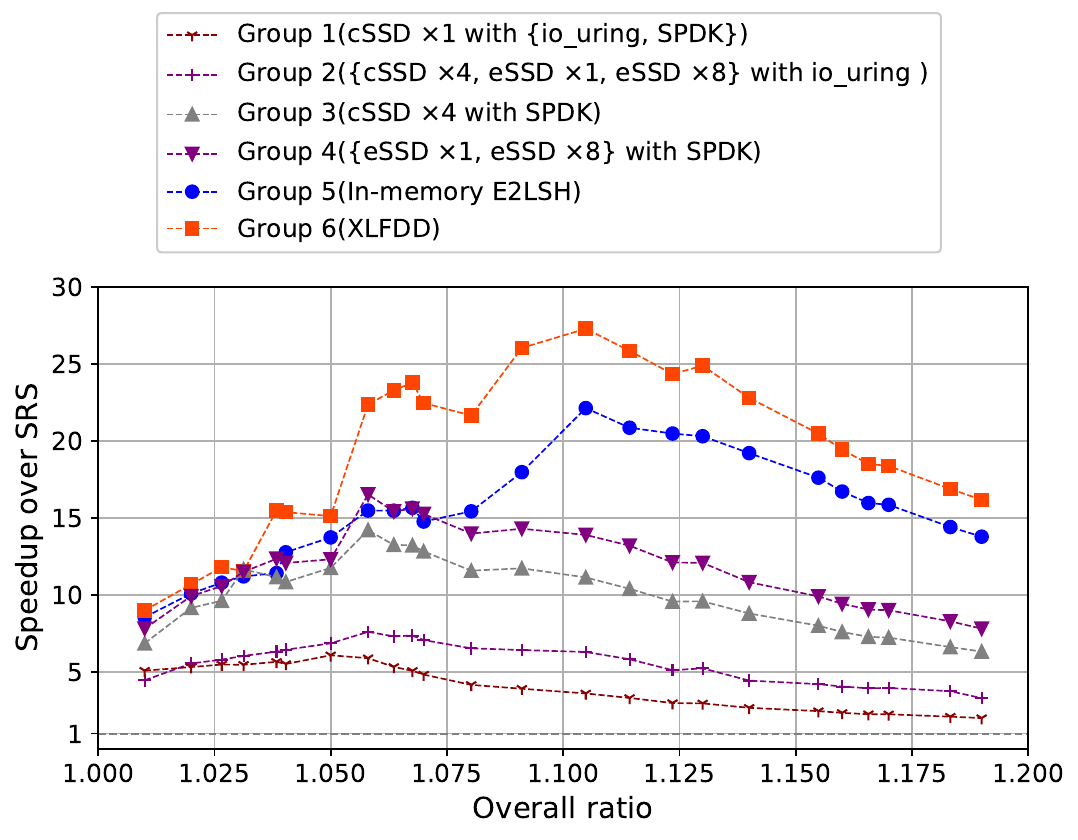}
  \vspace*{-7mm}
  \caption{Comparison of various storage configurations showing speedups over SRS for SIFT dataset}
  \label{fig:sift_vs_srs}
  \end{center}
\vspace{3mm}
  \begin{center}
  \includegraphics[width=0.8\columnwidth]{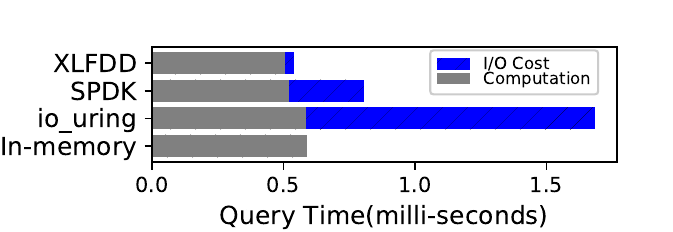}
  \caption{I/O cost of different storage interfaces for SIFT}
  \label{fig:latency_analysis}
  \end{center}
\end{figure}

We first investigate the impact of storage configurations using SIFT dataset.
Figure~\ref{fig:sift_vs_srs} shows speedups over SRS for various configurations.
As some storage configurations result in almost identical speedups, we represent them by a single line using their geometric mean to avoid clutter, leading to six distinct lines corresponding to six groups of configurations.
We explain them from bottom to top.
\begin{itemize}
\item {\bf Group 1 (cSSD $\times$ 1):}
The lowest line uses a single cSSD with either io\_uring or SPDK.
The speedup is above 1, meaning that E2LSHoS runs faster than SRS even with a single cSSD in agreement with our analysis in Sec~\ref{sec:srs_speed}.
The IOPS of a single cSSD limits the performance.

\item {\bf Group 2 (io\_uring):}
The next line is the result of $\rm cSSD \times 4$, $\rm eSSD \times 1$ and $\rm eSSD \times 8$, all using io\_uring.
All of them provide random read performance of over 1 MIOPS, indicating that the CPU overhead of io\_uring limits the performance.

\item {\bf Group 3 (cSSD $\times$ 4 with SPDK):}
The use of faster interface by SPDK boosts the performance over Group 2.

\item {\bf Group 4 (eSSD with SPDK):}
This line is the result of $\rm eSSD \times 1$ and $\rm eSSD \times 8$, both using SPDK.
As the in-memory-class IOPS requirement is slightly over what $\rm cSSD \times 4$ offers, eSSD gives another performance boost over Group 3.
As a single eSSD provides enough IOPS, using multiple of them does not lead to any further speed-up.

\item {\bf Group 5 (in-memory E2LSH):}
It is faster than E2LSHoS on eSSD with SPDK.
As eSSDs provide sufficient IOPS, the slowdown of E2LSHoS comes from the overhead of SPDK.

\item {\bf Group 6 (XLFDD):}
It reaches (and exceeds in this case) the in-memory speed with the lightweight interface of XLFDD.

\end{itemize}
These results show that both IOPS and CPU I/O overhead requirements must be met to achieve in-memory-class performance, in agreement with our analysis in Sec.~\ref{sec:in_memory_speed}.
The impact of storage interfaces can also be clearly seen in Figure~\ref{fig:latency_analysis}, comparing the I/O cost in the query time.
Here, $\rm eSSD \times 8$ is used with io\_uring and SPDK so IOPS will not be a limiting factor.
The I/O cost is 
the total CPU time spent for I/O-related functions measured with \textit{perf} command.

The reason why E2LSHoS on XLFDD outperforms in-memory E2LSH is attributed to the smaller memory footprint as explained in Sec.~\ref{sec:in_memory_speed}.
The reduced stall time overcompensates the I/O cost.


\subsection{Query Performance}

\begin{figure}[t]
  \begin{center}
  \includegraphics[width=1.0\linewidth]{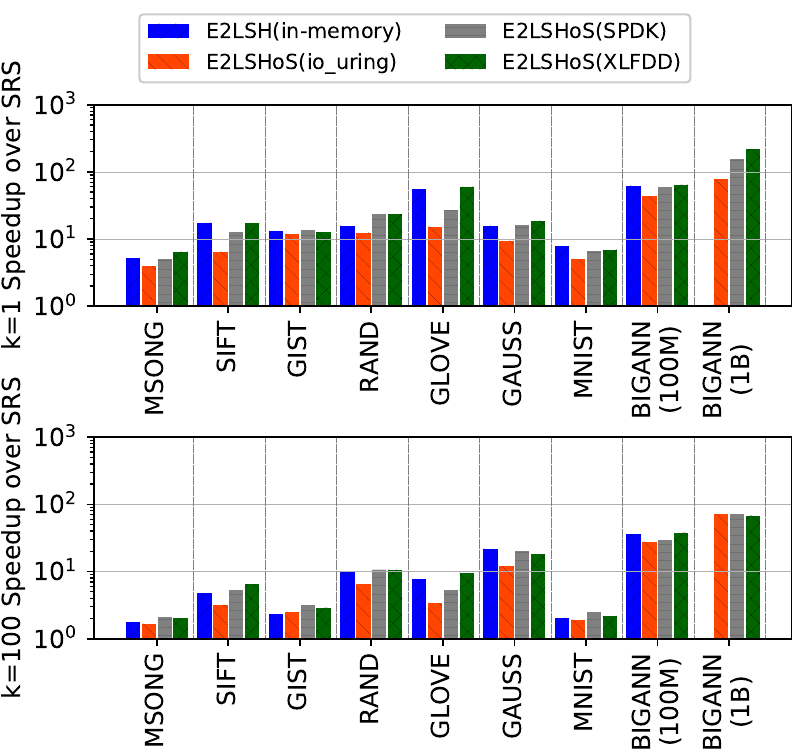}
  \vspace*{-3mm}
  \caption{Speedups over SRS for top-1 ANNS ($k=1$, top) and top-100 ANNS ($k=100$, bottom) at an overall ratio of 1.05}
  \label{fig:summary}
  \end{center}
\vspace{5mm}
\begin{center}
  \includegraphics[width=1.0\linewidth]{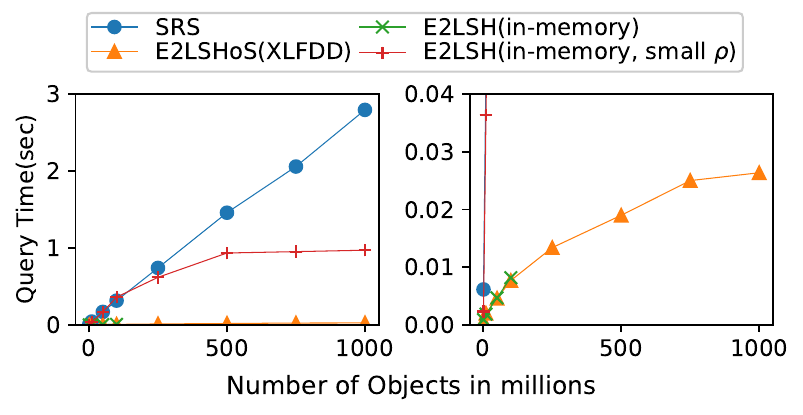}
  \vspace*{-7mm}
  \caption{Query times for varying database size $n$ at an overall ratio of 1.05 (two panels showing different time scales)}
  \label{fig:sublinear}
  \end{center}
\end{figure}

Next, we see if the tendency observed above in SIFT dataset holds in general for all the datasets.
We use the three storage interfaces as before, but
we limit the SSD configuration to $\rm cSSD \times 4$ as this represents a low-cost solution that still provides sufficient random read performance 
and capacity. 
Figure \ref{fig:summary} shows the speedup gains over SRS.
Note that the values of in-memory E2LSH for BIGANN dataset are missing because the DRAM capacity (768 GB)
of our machine is too small to run it with the same parameters as E2LSHoS.
A subset (100 million) of BIGANN is added for in-memory comparison.
The figure shows the impact of storage interfaces 
as observed in Sec.~\ref{sec:eval_storage_config}.
E2LSHoS approaches and sometimes exceeds in-memory E2LSH speeds using faster interfaces.
E2LSHoS consistently outperforms SRS. In particular, it is faster by an order of magnitude or two for BIGANN dataset.
The benefit of E2LSHoS becomes more significant for larger datasets thanks to its sublinear query time, which will be further evaluated below in Sec.~~\ref{sec:sublinear}.



\subsection{Validation of Sublinear Query Time}
\label{sec:sublinear}

Here we show that, as with in-memory E2LSH, the query time of E2LSHoS grows sublinearly with the database size $n$. 
To this end, we generate databases of varying sizes by taking subsets of BIGANN dataset.
We run E2LSHoS on XLFDDs and compare it with in-memory E2LSH and SRS.
For in-memory E2LSH, we conduct the experiment under two conditions: one with the same parameter $\rho$ with E2LSHoS, and the other with an extremely small $\rho = 0.09$ that dramatically reduces the index size to the extent that it permits in-memory execution for the one billion case.
Under the latter condition, the LSH cannot reliably find quality ANN candidates, which has to be compensated for by checking more candidates in order to achieve the target accuracy, significantly increasing the query time.
Figure \ref{fig:sublinear} shows the query times.
Since their ranges differ greatly, the two panels show the same plots with different vertical axis scales.
As can be seen in the left plots with a larger time scale, the SRS query time grows linearly and its difference from E2LSH(oS) becomes increasingly larger.
The right plots with a smaller scale demonstrates the sublinear growth of E2LSHoS.
Even though in-memory E2LSH follows the same curve as E2LSHoS when the same parameter is used, it has to stop at a database size of 100 million objects due to the DRAM capacity limit.
With the small $\rho$, in-memory E2LSH can reach one billion objects, but the query time becomes significantly longer than E2LSHoS.

\subsection{Index Size and Runtime Memory Usage}

Table~\ref{table:index_dram} compares E2LSHoS with SRS in terms of the index size and runtime memory usage.
As the database is placed on DRAM, the memory usage is the sum of the database size and the index size on memory (shown in parentheses) for both E2LSHoS and SRS.
While E2LSHoS uses a large index on storage, it only keeps relatively small index-related data (the hash table addresses) on DRAM, leading to comparable memory usage to SRS.

\begin{table}[h]
\small
  \caption{Index size and runtime memory usage}
  \label{table:index_dram}
 \centering
  \begin{tabular}{|l|r|rr|rr|}
    \hline
    & \multicolumn{3}{|c|}{E2LSHoS} & \multicolumn{2}{c|}{SRS} \\
    \cline{2-6}
     & Index   & Mem    & (Index & Mem   & (Index  \\
     & storage & usage  &  mem)  & usage &  mem)   \\
    \hline
    MSONG         &  4.6 GB &   1.7 GB & (3.6 MB) &   1.7 GB &  (47 MB) \\
    SIFT          &  6.3 GB &   133 MB & (3.9 MB) &   177 MB &  (48 MB) \\
    GIST          &  2.2 GB &   3.9 GB & (2.1 MB) &   3.9 GB &  (48 MB) \\
    RAND          &   10 GB &   412 MB & (8.2 MB) &   452 MB &  (48 MB) \\
    GLOVE         &   10 GB &   487 MB & (9.4 MB) &   534 MB &  (57 MB) \\
    GAUSS         &  6.0 GB &   4.1 GB & (4.9 MB) &   4.1 GB &  (96 MB) \\
    MNIST         &   16 GB &   6.4 GB & (6.6 MB) &   6.8 GB & (410 MB) \\
    BIGANN  &
    \multirow{2}{*}{530 GB} &
    \multirow{2}{*}{15 GB} &
    \multirow{2}{*}{(2.1 GB)} &
    \multirow{2}{*}{18 GB} &
    \multirow{2}{*}{(4.8 GB)} \\
    (100M)  &   &  &  & &  \\
    BIGANN &
    \multirow{2}{*}{6.1 TB} &
    \multirow{2}{*}{139 GB} &
    \multirow{2}{*}{(7.2 GB)} &
    \multirow{2}{*}{180 GB} &
    \multirow{2}{*}{(48 GB)} \\
    (1B)  &   &  &  & &  \\
    \hline
  \end{tabular}
\end{table}

\subsection{Additional Experiments}
\label{sec:additional_experiments}

{We present a few more experiments to further characterize the performance of E2LSHoS. Here, E2LSHoS on cSSDs uses io\_uring. 
}

\noindent
{\bf Comparison with synchronous I/Os:}
{In order to assess the impact of the asynchronous implementation of E2LSHoS, we compare its speed with that of a synchronous implementation.
To evaluate the latter, we run in-memory E2LSH with memory-mapped I/O, turning its DRAM accesses into storage I/Os.
With BIGANN(100M) dataset, we limit the page cache size to 32 GB, which is comparable to the E2LSHoS memory usage.
This synchronous implementation turns out 19.7 times slower than E2LSHoS (both use $\rm cSSD \times 4$).
This is because it does not hide the storage latency as illustrated in Figure~\ref{fig:io_interface}(A).
Even though we use the page cache, it is not effective.
In fact, the page cache miss rate (page fault per 4 kB read) is as high as 93\% due to the random access nature of the E2LSH algorithm.
}

\begin{figure}[t]
  \begin{center}
  \includegraphics[width=1.0\linewidth]{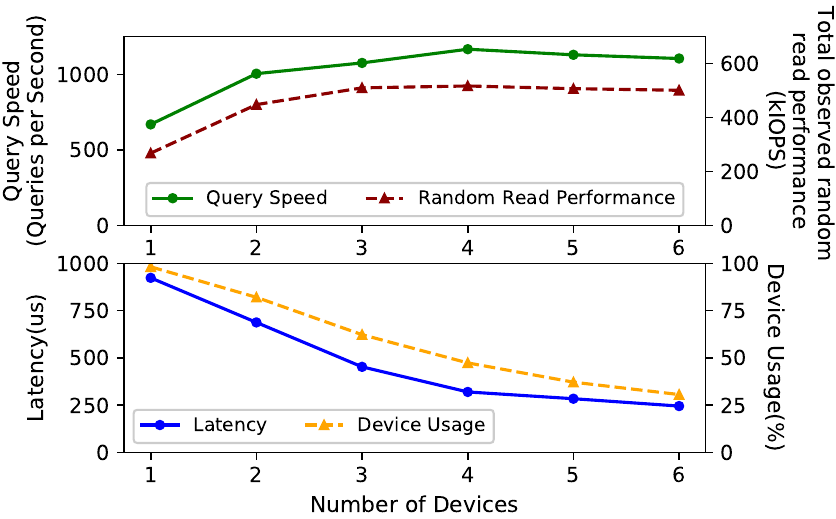}
  \vspace*{-6mm}
  \caption{Query speeds and device statistics for varying number of devices}
  \label{fig:num_devices}
  \end{center}
\vspace{5mm}
  \begin{center}
  \includegraphics[scale=0.6]{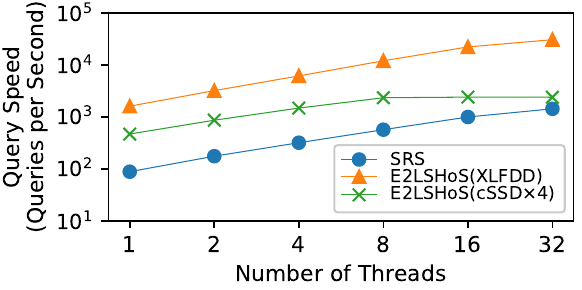}
  \vspace*{-3mm}
  \caption{Query speeds with multithreading}
  \label{fig:multithread}
  \end{center}
\end{figure}

\noindent
{\bf Storage throughput and latency:}
{When storage devices run at a high IOPS, their latency becomes long.
In order to examine how this trade-off affects the query speed of E2LSHoS,
we vary the number of cSSDs for SIFT dataset
and plot the query speed (in queries per second), total observed IOPS, latency, and device usage (per-device observed IOPS divided by the maximum random read performance, which is 273 kIOPS) in Figure~\ref{fig:num_devices}.
The result indicates that the query speed is proportional to the IOPS value, supporting our analysis that the random read performance is the primary storage metric determining the query speed.
The speed increases as we add devices until the maximum total IOPS the devices can sustain exceeds what the workload requires.
When the number of devices is small and their usage is high, the latency becomes longer, but the latency by itself does not determine the application performance.
}

\noindent
{\bf Multithreading:}
{All of our evaluations so far have been done on a single thread.
We evaluate the query speeds of E2LSHoS and SRS using up to 32 threads for SIFT dataset as shown in Figure~\ref{fig:multithread}.
Both methods linearly scale, except that E2LSHoS plateaus when it is bottlenecked by the storage IOPS.
E2LSHoS on cSSDs is faster than SRS for up to 8 threads, but the gap narrows beyond it.
E2LSHoS on XLFDDs is consistently faster than SRS by an order of magnitude.
}

\section{Discussion}
\label{sec:discussion}

We believe our analysis and evaluation have demonstrated that E2LSHoS can be a viable alternative to small-index LSH methods, 
but there are a number of factors to be considered 
in practice.

\noindent
{\bf Storage capacity and cost:}
E2LSHoS runs fast thanks to its sublinear time complexity, but its price remains to be the large hash index size.
{It does not reduce the space complexity of E2LSH, and its large space consumption translates to additional hardware cost, which is a major limitation of this approach.}
Nonetheless, with E2LSHoS, the limit of E2LSH has been greatly relaxed, increasing the maximum allowable index size and enabling the algorithm to run even with one billion objects.
However, depending on the system and problem at hand, the desired accuracy and speed may not be achieved due to the storage capacity limit.
While we are hopeful that the continued exponential growth trend of storage capacity will render this less of a concern, it would be interesting to consider incorporating the ideas from small-index methods in such a way that the index size of E2LSHoS is reduced without sacrificing its sublinear query time.

{Compared with small-index methods, the cost increase due to storage may be justified by the application performance gain. For one-billion ANNS, we added 6 TB of SSDs (around \$2,000) to our \$15,000 server. Therefore, with 13\% more hardware cost, we were able to achieve 100 times faster query than SRS.
}

\noindent
{\bf Storage-specific issues:}
{
Using storage requires attention to the associated issues including endurance and heat.
As SSDs have a limit to the amount of data that can be written under warranty, updating the hash index consumes the device life.
While the impact of object insertion and deletion is small, rebuilding the entire index should be done sparingly.
%
Under heavy load, SSDs may trigger thermal throttling to prevent overheating, causing temporary performance degradation. Adequate cooling is important to avoid this.}

\noindent
{\bf More general implementations:}
{
Our approach is not the only way of executing E2LSH on external memory.
More general implemetations would be to develop a mechanism that would mediate in-memory E2LSH and storage so that it would serve DRAM loads while fetching data from storage.
Memory-mapped I/O with the page cache in Sec.~\ref{sec:additional_experiments} can be viewed as one such implementation, albeit slow: storage latency and CPU I/O overhead including the page cache account for 50\% and 40\% of the query time, respectively.
Therefore, an efficient mediation mechanism would need not only to employ a lightweight user-space cache, but also to hide the storage latency to cope with the cache-unfriendly E2LSH workload in such a way that handling cache misses (including resultant I/Os) could be overlapped with computation using many threads and contexts.
While we have taken conceptually similar strategies for our E2LSHoS implementation in an application-specific way, designing a more general mechanism is an interesting future direction.}

\noindent
{\bf Non-storage solutions:}
We have shown that large databases requiring a terabyte-sized index can be handled by E2LSHoS on a single node.
However, we are not trying to imply that single-node in-memory E2LSH is infeasible for large databases.
Some flagship machines can accommodate 24 terabytes of DRAM per node \cite{Lenovo,AWS}.
Moreover, storage class memory such as Intel\textregistered{} Optane\texttrademark{} DC Persistent Memory Module may be used to expand the system main memory further, for instance to 36 terabytes \cite{Lenovo}.
Therefore, it is up to the user what environment to run E2LSH(oS) on.
We believe E2LSHoS offers a lower-cost solution to sublinear time ANNS.

\section{Conclusion}



This paper has shown that E2LSH 
is regaining the advantage in query speed over small-index LSH methods 
with the advent of modern flash storage devices. 
We have 
analyzed the E2LSH algorithm on a modern single-node computing environment, and shown that (1) the storage performance necessary for E2LSHoS to run faster than small-index methods is satisfied by a single consumer-grade NVMe\texttrademark{} SSD,
and (2) emerging high-performance storage devices and interfaces allow E2LSHoS to approach in-memory E2LSH speeds.
Our E2LSHoS implementation has demonstrated these two points for large datasets of up to one billion objects with comparable DRAM usage to small-index methods.
These results indicate that the user can enjoy the benefit of sublinear query time of the E2LSH algorithm beyond the index size limit of in-memory E2LSH.
%

We believe our work suggests that
large-index LSH methods 
are becoming increasingly worth exploring.
{In particular, methods using similar index structures to E2LSH, such as \cite{LSBFOREST,MULTIPROBELSH}, are likely to benefit from modern storage devices.}
However, we do not intend to imply large-index methods are superior. 
The benefits of small-index methods are clear and they will continue to be valuable.
{Moreover, small-index methods may also benefit from modern storage devices on a memory-limited environment.
Asynchronous I/Os will be useful in such cases as well.
For example, external-memory SRS and QALSH may issue requests for adjacent tree nodes while processing the current node.}
In summary, our intention is to renew the community's interest in larger-index methods as well as to spur efforts of leveraging modern storage devices.
We believe this will help expand the possible solution space of LSH-based ANNS,
from which we hope further new approaches will emerge.

\bibliographystyle{ACM-Reference-Format}
\bibliography{mybibfile}

\end{document}